# A comparison of university performance scores and ranks by MNCS and FSS[1]

*Giovanni Abramo* (corresponding author)
    Laboratory for Studies of Research and Technology Transfer
    Institute for System Analysis and Computer Science (IASI-CNR)
    National Research Council of Italy
    Via dei Taurini 19, 00185 Rome - ITALY
    giovanni.abramo@uniroma2.it
    Tel/fax +39 06 72597362

*Ciriaco Andrea D'Angelo*
    Department of Engineering and Management
    University of Rome "Tor Vergata"
    Via del Politecnico 1, 00133 Rome - ITALY
    dangelo@dii.uniroma2.it

**Abstract**

In a previous article of ours, we explained the reasons why the MNCS and all similar per-publication citation indicators should not be used to measure research performance, whereas efficiency indicators (output to input) such as the FSS are valid indicators of performance. The problem frequently indicated in measuring efficiency indicators lies in the availability of input data. If we accept that such data are inaccessible, and instead resort to per-publication citation indicators, the question arises as to what extent institution performance rankings by MNCS are different from those by FSS (and so what effects such results could have on policy-makers, managers and other users of the rankings). Contrasting the 2008-2012 performance by MNCS and FSS of Italian universities in the Sciences, we try to answer that question at field, discipline, and overall university level. We present the descriptive statistics of the shifts in rank, and the correlations of both scores and ranks. The analysis reveals strong correlations in many fields but weak correlations in others. The extent of rank shifts is never negligible: a number of universities shift from top to non-top quartile ranks.

# 1. Introduction

The increasing application of New Public Management to the academic sector, with emphasis on quasi-market competition, efficiency and performance audit practices (Schubert, 2009), has led to a situation of an influential and growing community of scientometricians, engaged in intense search for ever better research performance indicators. In recent years there has been a proliferation of new indicators and variants or extensions of old ones. At the same time, we witnessed a surge of international and national research performance rankings. These are based on different indicators and methods, and seem often to show contrasting results. While the ostensible aim was to support the policy makers and the managers of research institutions in making evidence-based decisions, the outcome is that of bewilderment: often the policy maker cannot discern the appropriate and valid methods to measure research performance, or the ranking on which to base their decisions. In our view, the moment has arrived for scientometricians to take responsibility; to converge on a synthesis stating which are the more appropriate indicators of performance.

In a recent work of ours (Abramo and D'Angelo, 2016a), we try to explain why the world-famous Leiden group's Mean Normalized Citation Score or MNCS (Waltman et al. 2011), cannot be considered a "performance" indicator and, therefore, the university rankings by MNCS are not valid. In the same special section we responded to the comments of eminent scholars in the field, and further argued our position on the matter (Abramo and D'Angelo, 2016b). In short, to us all size-independent indicators based on the ratio to publications are invalid indicators of performance, because research performance cannot be defined as the average impact of output (MNCS). Furthermore, performance (as measured by MNCS) may actually diminish if additional output is cited below average, which is a paradox. Vice versa, size-independent indicators based on the ratio to research input, are more appropriate indicators of performance, since they establish which individuals or research units, under parity of resources, have more or less impact on scientific advancement. Since the very beginning of our research activity in the field of scientometrics (Abramo et al., 2008a), we have always refrained from the adoption of such well established and already popular indicators as the $h$-index (Hirsch, 2005) and the CPP/FCSm or "old" crown indicator (Van Raan, 2005; Moed et al., 1995), the forerunner of the current MNCS. Instead, we pursued the measurement of efficiency indicators which could allow the ranking of individuals and research units according to a better proxy of their "real" performance, despite all the assumptions and limits embedded in the operationalization of the measurement. The latest versions and the detailed explications of the theory underlying the two indicators that we apply to approximate the measure of labor productivity in research institutions, namely the *Fractional Scientific Strength* (FSS) and the HCAs (highly cited articles) per scientist, can be found in Abramo and D'Angelo (2014) and Abramo and D'Angelo (2015a).

The limits of the $h$-index have been discussed extensively in the literature and there have been numerous attempts to overcome them through $h$-variants (Egghe, 2010; Norris and Oppenheim, 2010; Alonso et al., 2009). In two previous works of ours, we have measured the differences in university rankings by FSS and h- and g-indexes (Abramo et al., 2013a), as well as at the individual level (Abramo et al., 2013b). In this work we intend to do the same for the MNCS, to see to what extent the university performance scores and ranks by FSS diverge from those by MNCS. We will assess the differences at field, discipline and overall institution level.

By common sense one would expect that in general talented researchers capable to produce high impact publications do also produce a high number of articles. Whereas less talented researchers produce a lower number of publications of lower impact. Leaving aside the few exceptions that prove the rule, several empirical studies confirm that. Abramo, D'Angelo & Di Costa (2010) demonstrate the existence of a strong correlation between quantity and impact of research production: scientists that are more productive in terms of quantity also achieve higher levels for impact in their research products. Larivière & Costas (2015) show that the higher the number of papers a researcher publishes, the more likely they are amongst the most cited in their domain. van den Besselaar & Sandström (2015) show that researchers producing a high number of papers have significantly higher probability to produce top cited papers. Since FSS embeds both quantity and impact of publications, because of the strong correlation between the two, one would expect a strong correlation between performance scores and ranks by FSS and MNCS. Our findings show that this is more or less true at discipline and at the aggregate institution level, although cases of noticeable shifts in ranking are registered.

The manuscript proceeds as follows: in the next section we present the field of observation and methodology adopted; Section 3 reports the results of the comparison; Section 4 provides the conclusions.

## 2. Data and Methods

### 2.1 Dataset

The dataset of the analysis is based on the 2008-2012 WoS indexed publications authored by professors in the Sciences of all Italian universities. Citations are observed at October, 2015. The Italian Ministry of Education, Universities and Research (MIUR) recognizes a total of 96 universities authorized to grant legally recognized degrees. In Italy there are no "teaching-only" universities, as all professors are required to carry out both research and teaching, in keeping with the Humboldtian philosophy of higher education. Each professor is officially classified in one and only one research field. There are a total of 370 such fields (named scientific disciplinary sectors, or SDS[2]), grouped into 14 disciplines (named university disciplinary areas, or UDAs). For reasons of robustness, we examine only the nine UDAs in the Sciences[3], including a total of 192 SDSs, whereby publications in indexed journals is the prevalent mode for output codification. Furthermore, again for robustness, we exclude all professors who have been on staff less than three years in the observed period (Abramo et al., 2012a).

Data on academics are extracted from a database maintained at the central level by the MIUR,[4] indexing the name, academic rank, affiliation, and the SDS of each professor. Publication data are drawn from the Italian Observatory of Public Research (ORP), a database developed and maintained by the authors and derived under license from the Web of Science (WoS). Beginning from the raw data of Italian publications[5] indexed in WoS-ORP, we apply an algorithm for disambiguation of the true identity of the authors and their institutional affiliations (for details see D'Angelo et al., 2011). Each publication is attributed to the university professors that authored it, with a harmonic average of precision and recall (F-measure) equal to 97 (error of 3%). We further reduce this error by manual disambiguation.

---

[2] The complete list is on http://attiministeriali.miur.it/UserFiles/115.htm, last accessed 05/07/2016.
[3] Mathematics and computer sciences; Physics; Chemistry; Earth sciences; Biology; Medicine; Agricultural and veterinary sciences; Civil engineering; Industrial and information engineering.
[4] http://cercauniversita.cineca.it/php5/docenti/cerca.php, last accessed 05/07/2016.

The dataset for the analysis includes 36,450 professors, employed in 86 universities, authoring over 200,000 WoS publications, sorted in the UDAs as shown in Table 1.

*Table 1: Dataset for the analysis. Number of fields (SDSs), universities, professors and WoS publications (2008-2012) in each UDA under investigation*

| UDA | SDS | Universities | Professors | Publications* |
|---|---|---|---|---|
| Mathematics and computer science | 10 | 69 | 3,387 | 16,920 |
| Physics | 8 | 64 | 2,497 | 23,587 |
| Chemistry | 12 | 61 | 3,174 | 26,703 |
| Earth sciences | 12 | 47 | 1,199 | 6,148 |
| Biology | 19 | 66 | 5,198 | 34,399 |
| Medicine | 50 | 64 | 10,966 | 71,575 |
| Agricultural and veterinary sciences | 30 | 55 | 3,207 | 14,209 |
| Civil engineering | 9 | 53 | 1,583 | 6,908 |
| Industrial and information engineering | 42 | 73 | 5,239 | 40,246 |
| Total | 192 | 86 | 36,450 | 206,433[†] |

\* The figure refers to publications (2008-2012) authored by at least one professor pertaining to the UDA.
[†] The total is less than the sum of the column data due to multiple counting of individual publications that pertain to the SDSs of more than one UDA.

**2.2 Measuring research performance by FSS and MNCS**

The MNCS and FSS are both impact indicators; as well, they are both size-independent indicators, meaning that the results are independent of the size of the institutions.[6] Very simply, the main conceptual difference between the two is that the former measures the average impact of the publications of a research unit, and the latter the average impact of the researchers. The MNCS belongs to a type of "per publication" impact indicator, while the FSS is an efficiency indicator.

The performance of an institution by FSS is based on the measurement of the individual performance of each scientist on staff. This measurement requires data on the scientists on staff in each institution, and the disambiguation of the authors' names for all publications. To this purpose, D'Angelo et al. (2011) have developed a disambiguation algorithm applicable to all professors of Italian universities. To date, the FSS has been used to rank the performance of only Italian institutions (meaning also the individual researchers). Compared to the FSS, the measurement of the MNCS is a simpler undertaking, requiring only the reconciliation of the names of the institutions. It has therefore been possible to apply the MNCS to ranking of institutions at the worldwide level.

MNCS and FSS adopt the fractional counting method[7] and field-normalize the citations to account for different citation behavior among fields. Although conceptually aligned on the above two features, notable differences occur when it comes to operationalizing the counting and normalization.

---

[5] We exclude those document types that cannot be strictly considered as true research products, such as editorial material, meeting abstracts, replies to letters, etc.
[6] If we exclude potential returns to size in research activity, as confirmed in the literature (Abramo et al., 2012b; Bonaccorsi and Daraio, 2005; Seglen and Asknes, 2000; Golden and Carstensen, 1992).
[7] Actually, at CWTS they also measure the MNCS using the full counting method.

For instance, the Leiden world rankings by MNCS adopt an address-level fractional counting, which fractionalizes publications by the number of address lines. Using our authorship disambiguation algorithm, we adopt an author-level fractional counting. Moreover, in the FSS the fractional contribution equals the inverse of the number of authors in those fields where the practice is to place the authors in simple alphabetical order, while in other cases it weights each contribution. For the life sciences, widespread practice in Italy is for the authors to indicate the various contributions to the published research by the order of the names in the byline. For the life science then, the FSS gives different weights to each co-author according to their position in the list of authors (Abramo et al. 2013c). Furthermore, being an efficiency indicator, the FSS also normalizes by the salary of each professor, to avoid favoring universities with a higher proportion of higher academic ranks, which therefore have a higher average cost per unit of labor (Abramo et al., 2010).

In the FSS, the citations of a publication $i$ are normalized to the average of the distribution of citations received for all Italian cited publications indexed in the same year and field as the publication $i$.[8] Differently, in the MNCS citations are normalized to the average of the distribution of citations received for all world publications, not just Italian and cited ones. Furthermore, the fields equal the 251 WoS subject category in the FSS; while in the MNCS, normalization is sometimes based on the WoS subject categories and sometimes on about 4000 fields. These fields are defined at the level of individual publications. Using a computer algorithm, the CWTS group assigns each WoS publication to a field based on its citation relations with other publications.[9]

Since the MNCS averages the citations per publication, it should be little affected[10] by the different intensity of publication across fields (Abramo and D'Angelo, 2015b; Butler, 2007; Moed et al., 1985; Garfield, 1979). Instead, since the FSS calculates the total impact, it is indeed affected. In fact, all else being equal, the higher the number of cited publications the higher the FSS. When applying the FSS then, in order to avoid distortions (Abramo et al., 2008b), the researchers must be classified in their respective fields, with their performance then normalized by a field-specific scaling factor.

Because our intent here is to assess the differences in scores and ranks by MNCS and FSS, caused by the different conceptualization of the two indicators, rather than by the operationalization of the measure, we have aligned as much as possible the fractional counting and the field-normalization methods. For both indicators, we measure the fractional contribution as the inverse of the number of authors, without weighting it according to the position of the authors in the byline. The fields to normalize citations are the WoS subject categories[11]. The chosen scaling factor is the average citations received for all Italian cited publications. The reason for this choice is that we reckon it more appropriate when it comes to compare institutions within the same country. It avoids favoring institutions carrying out research in fields where the country is on the frontier, vis-à-vis institutions carrying out catch-up research in fields where the country lags.

---

[8] Abramo et al. (2012c) demonstrated that the average of the distribution of citations received for all cited publications of the same year and subject category is the most effective scaling factor.
[9] For an explanation of the procedure, see Ruiz-Castillo & Waltman (2015).
[10] This expectation could be a subject worthy of further investigation.
[11] The subject category of a publication corresponds to that of the journal where it is published. For publications in journals belonging to more than one category, the scaling factor is calculated as the average of the scaling factors for each subject category.

Following are the formulae of FSS and MNCS applied in this work.

At the level of the individual professor *P*, the average yearly productivity FSS, accounting for the cost of labor, is:

$$FSS_P = \frac{1}{w_P} \cdot \frac{1}{t} \sum_{i=1}^{N} \frac{c_i}{\bar{c}} \cdot \frac{1}{n_i}$$

[1]

Where:
$w_P$ = average yearly salary of the professor;[12]
t = number of years the professor worked over the period of observation;
N = number of publications by the professor over the period of observation;
$c_i$ = citations received for publication *i*;
$\bar{c}$ = average of the distribution of citations received for all Italian cited publications indexed in the same year and subject category as publication *i*;
$n_i$ = number of all co-authors (including non Italian) of publication *i*.

University productivity in a field, discipline or "overall" involves standardization of individual productivity by the SDS average (Abramo & D'Angelo, 2015). In formula, the productivity $FSS_U$ over a certain period for university *U*, in a field, discipline and overall is:

$$FSS_U = \frac{1}{RS} \sum_{j=1}^{RS} \frac{FSS_j}{\overline{FSS}}$$

[2]

Where:
$RS$ = research staff of the field/discipline/university, in the observed period;
$FSS_j$ = productivity of professor *j*;
$\overline{FSS}$ = national average productivity of all productive professors in the same SDS as professor *j*.

The reader is referred to Abramo and D'Angelo (2014) for a more detailed explication of the theory underlying this indicator.

For a generic university, the MNCS is measured here as follows:

$$MNCS = \frac{\sum_{i=1}^{M} \frac{c_i}{\bar{c}} \cdot \frac{m_i}{n_i}}{\sum_{i=1}^{M} \frac{m_i}{n_i}}$$

[3]

Where
M = number of publications by the university over the period of observation;
$c_i$ = citations received for publication *i*;
$\bar{c}$ = average of the distribution of citations received for all Italian cited publications indexed in the same year and subject category as publication *i*;

---

[12] This information is unavailable for reasons of privacy. We resort to a proxy, i.e. the nationally averaged salary of the professors in each academic rank (data source DALIA – MIUR, https://dalia.cineca.it/php4/inizio_access_cnvsu.php, last accessed 05/07/2016). Failure to account for the cost of labor would result in ranking distortions, as shown by Abramo et al. (2010).

$m_i$ = number of co-authors of university of publication *i*,
$n_i$ = total number of co-authors (including non Italian ones) of publication *i*.

## 3. Results

In this section we present the results of the comparisons of the university performance scores and ranks by FSS and MNCS obtained by [2] and [3], at SDS, UDA and overall university level.

### 3.1 Comparing university scores and rankings at the field level

To carry out the comparison, we have measured the performance scores and ranks by FSS and MNCS of all universities in each SDS. We exclude those universities with less than two professors in the SDS. To exemplify, we present the case of the SDSs of Chemistry (UDA 3). Table 2 shows the comparison of scores and rankings in the SDS Pharmaceutical chemistry (CHIM/08). In this SDS, 29 universities have more than two professors. The maximum negative percentile shift is -35.7 (10 positions), while the maximum positive one is +50.0 (14 positions gained in the ranking). To better appreciate the entities of the shifts, Figure 1 presents a graphic view of the dispersion of FSS and MNCS scores, while Figure 2 shows the percentiles and percentile rank differences registered for each university. The correlation between the scores by the two indicators (Pearson $\rho$) is 0.864, the rank correlation (Spearman $\rho$) is 0.756. Because of space limits, we cannot present in detail the results for all 9 UDAs and 192 SDSs. To sustain our arguments, we could have chosen to show those disciplines and fields where the score and rank correlations between the two indicators is weak, but the attentive reader realizes that differences can only be larger in other disciplines. For example, Figure 3 presents the dispersion of scores for the 51 universities in FIS/01, Experimental physics, where the correlations are weak (Pearson $\rho$ = 0.326; Spearman $\rho$ = 0.400).

*Table 2: Comparison of scores and rankings by FSS and MNCS for Italian universities in CHIM/08 (Pharmaceutical Chemistry)*

| ID* | Research staff | FSS | | | MNCS | | | Rank shift | Percentile shift |
|---|---|---|---|---|---|---|---|---|---|
| | | score | rank | percentile | score | rank | percentile | | |
| UNIV_1 | 22 | 4.035 | 1 | 100.0 | 2.158 | 1 | 100 | = | 0.0 |
| UNIV_2 | 3 | 1.581 | 2 | 96.4 | 1.199 | 5 | 86 | ↓3 | -10.7 |
| UNIV_3 | 34 | 1.441 | 3 | 92.9 | 1.321 | 3 | 93 | = | 0.0 |
| UNIV_4 | 14 | 1.175 | 5 | 85.7 | 0.889 | 14 | 54 | ↓9 | -32.1 |
| UNIV_5 | 12 | 0.886 | 11 | 64.3 | 0.855 | 15 | 50 | ↓4 | -14.3 |
| UNIV_6 | 14 | 1.220 | 4 | 89.3 | 1.065 | 8 | 75 | ↓4 | -14.3 |
| UNIV_7 | 7 | 1.140 | 6 | 82.1 | 1.450 | 2 | 96 | ↑4 | +14.3 |
| UNIV_8 | 13 | 0.785 | 14 | 53.6 | 0.953 | 10 | 68 | ↑4 | +14.3 |
| UNIV_9 | 21 | 0.970 | 10 | 67.9 | 0.916 | 12 | 61 | ↓2 | -7.1 |
| UNIV_10 | 21 | 1.078 | 7 | 78.6 | 0.822 | 17 | 43 | ↓10 | -35.7 |
| UNIV_11 | 26 | 0.974 | 9 | 71.4 | 1.082 | 6 | 82 | ↑3 | +10.7 |
| UNIV_12 | 14 | 1.000 | 8 | 75.0 | 0.927 | 11 | 64 | ↓3 | -10.7 |
| UNIV_13 | 25 | 0.871 | 12 | 60.7 | 0.774 | 21 | 29 | ↓9 | -32.1 |
| UNIV_14 | 10 | 0.661 | 21 | 28.6 | 0.806 | 19 | 36 | ↑2 | +7.1 |
| UNIV_15 | 33 | 0.778 | 15 | 50.0 | 0.793 | 20 | 32 | ↓5 | -17.9 |
| UNIV_16 | 10 | 0.720 | 19 | 35.7 | 0.912 | 13 | 57 | ↑6 | +21.4 |
| UNIV_17 | 24 | 0.794 | 13 | 57.1 | 1.004 | 9 | 71 | ↑4 | +14.3 |
| UNIV_18 | 10 | 0.702 | 20 | 32.1 | 1.082 | 7 | 79 | ↑13 | +46.4 |
| UNIV_19 | 14 | 0.752 | 16 | 46.4 | 0.698 | 24 | 18 | ↓8 | -28.6 |
| UNIV_20 | 12 | 0.732 | 17 | 42.9 | 0.820 | 18 | 39 | ↓1 | -3.6 |
| UNIV_21 | 8 | 0.726 | 18 | 39.3 | 1.313 | 4 | 89 | ↑14 | +50.0 |
| UNIV_22 | 10 | 0.588 | 23 | 21.4 | 0.845 | 16 | 46 | ↑7 | +25.0 |
| UNIV_23 | 12 | 0.510 | 24 | 17.9 | 0.580 | 27 | 7 | ↓3 | -10.7 |
| UNIV_24 | 5 | 0.615 | 22 | 25.0 | 0.677 | 25 | 14 | ↓3 | -10.7 |
| UNIV_25 | 16 | 0.442 | 28 | 3.6 | 0.760 | 22 | 25 | ↑6 | +21.4 |
| UNIV_26 | 15 | 0.510 | 25 | 14.3 | 0.618 | 26 | 11 | ↓1 | -3.6 |
| UNIV_27 | 24 | 0.448 | 27 | 7.1 | 0.699 | 23 | 21 | ↑4 | +14.3 |
| UNIV_28 | 21 | 0.479 | 26 | 10.7 | 0.580 | 28 | 4 | ↓2 | -7.1 |
| UNIV_29 | 19 | 0.274 | 29 | 0.0 | 0.503 | 29 | 0 | = | 0.0 |

* The population consists of universities (29 in all) having at least two professors in the SDS

*Figure 1: FSS and MNCS scores of Italian universities in CHIM/08 (Pharmaceutical chemistry)*

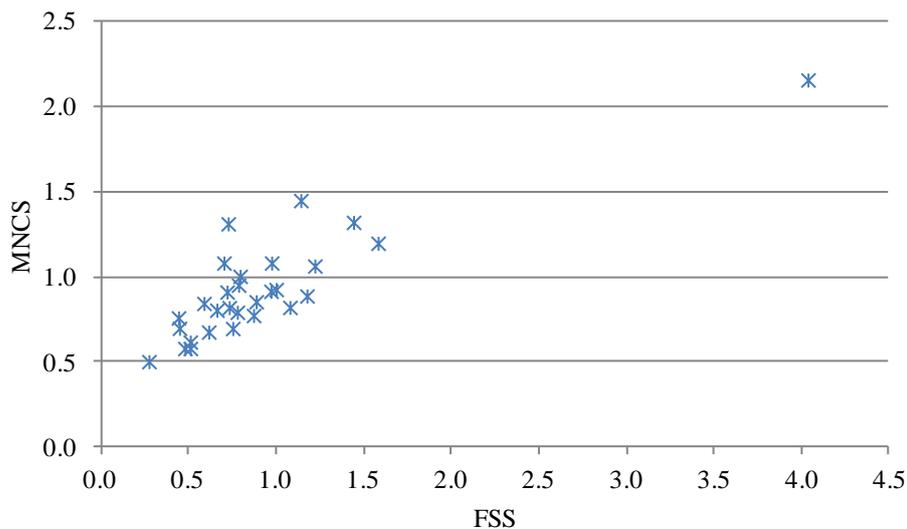

*Figure 2: University rankings (percentile) by FSS and MNCS in CHIM/08 (Pharmaceutical Chemistry)*

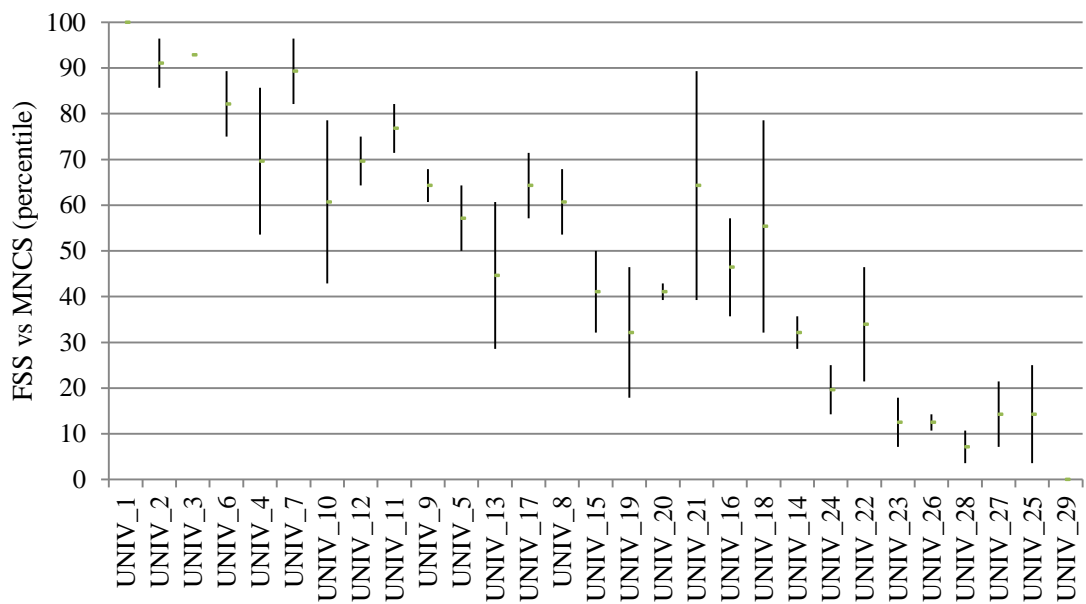

*Figure 3: FSS and MNCS scores of Italian universities in FIS/01 (Experimental physics)*

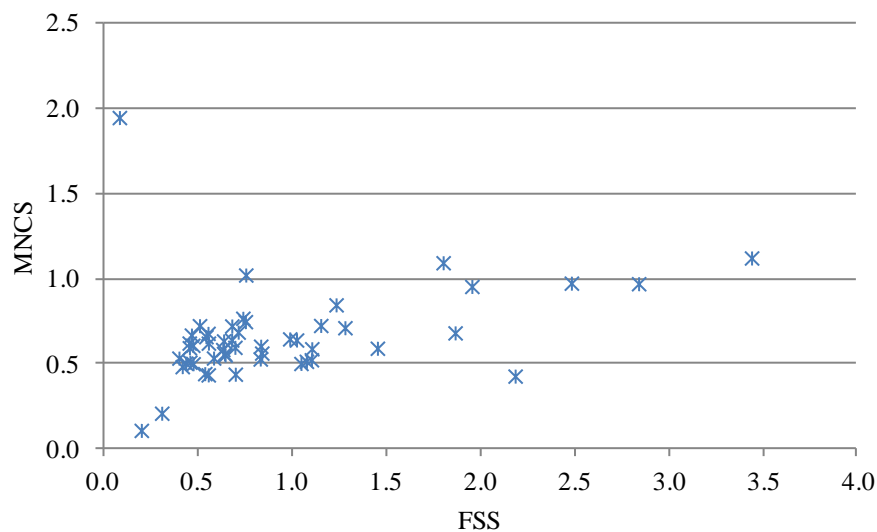

Table 3 presents the descriptive statistics of university ranking shifts by FSS and MNCS in each SDS of Chemistry.[13] The correlation is strong in all SDSs but CHIM/11 and CHIM/07, although a very high number of universities experience a shift in ranking in each SDS and the maximum shifts in each SDS are clearly notable (in CHIM/07 one university experience a shift of 28 positions out of 31).

---

[13] CHIM/05 has been excluded because only five professors belong to it.

*Table 3: Score and rank correlations, and descriptive statistics of university ranking shifts by FSS and MNCS in the SDSs of Chemistry (percentile shift in brackets)*

| SSD* | Universities | Pearson ρ | Spearman ρ | % shifting rank | Average shift | Median shift | Max shift |
|---|---|---|---|---|---|---|---|
| CHIM/01 | 39 | 0.782 | 0.806 | 92.3% | 4.9 (12.8) | 3 | 24.0 (63.2) |
| CHIM/02 | 36 | 0.655 | 0.849 | 88.9% | 4.3 (12.2) | 4 | 21.0 (60.0) |
| CHIM/03 | 42 | 0.660 | 0.891 | 90.5% | 4.0 (9.9) | 3 | 16.0 (39.0) |
| CHIM/04 | 18 | 0.868 | 0.959 | 77.8% | 1.2 (7.2) | 1 | 3.0 (17.6) |
| CHIM/06 | 45 | 0.561 | 0.699 | 91.1% | 7.4 (16.8) | 5 | 33.0 (75.0) |
| CHIM/07 | 31 | 0.437 | 0.586 | 93.5% | 5.7 (19.1) | 5 | 28.0 (93.3) |
| CHIM/08 | 29 | 0.864 | 0.756 | 89.7% | 4.6 (16.5) | 4 | 14.0 (50.0) |
| CHIM/09 | 28 | 0.645 | 0.772 | 82.1% | 3.8 (14.0) | 2.5 | 13.0 (48.1) |
| CHIM/10 | 18 | 0.807 | 0.750 | 83.3% | 2.8 (16.3) | 2 | 8.0 (47.1) |
| CHIM/11 | 8 | -0.173 | 0.000 | 100.0% | 2.8 (39.3) | 2 | 6.0 (85.7) |
| CHIM/12 | 15 | 0.665 | 0.514 | 86.7% | 2.7 (19.0) | 1 | 11.0 (78.6) |

* CHIM/01 = Analytical Chemistry; CHIM/02 = Physical Chemistry; CHIM/03 = General and Inorganic Chemistry; CHIM/04 = Industrial Chemistry; CHIM/06 = Organic chemistry; CHIM/07 = Foundations of Chemistry for Technologies; CHIM/08 = Pharmaceutical Chemistry; CHIM/09 = Applied Technological Pharmaceutics; CHIM/10 = Food Chemistry; CHIM/11 = Chemistry and Biotechnology of Fermentations; CHIM/12 = Environmental Chemistry and Chemistry for Cultural Heritage

In Table 4 we present the descriptive statistics of university ranking shifts by FSS and MNCS in the SDSs of all UDAs. Except for UDA 9, the number of universities experiencing shifts is never below 46.2% of the total in an SDS, but is as high as 100% in several SDSs. The maximum shift in ranking is never below 17.6, furthermore in Physics (UDA 2), Medicine (UDA 6) and Agricultural and veterinary sciences (UDA 7) there is at least one SDS where the top university shifts to bottom or vice versa. UDA 9, presents the case of an SDS where the two rankings are identical. Pearson and Spearman correlations between the two rankings is extremely strong in a number of SDSs, but is also very weak in others.

*Table 4: Comparison of university rankings by FSS and MNCS (min-max of percentile variations) and correlations for the SDSs of each UDA*

| UDA* | No. of SDSs** | Range of variation (min-max) of universities experiencing shift (%) | Range of variation (min-max) of average shift (percentiles) | Range of variation (min-max) in max shift (percentiles) | Pearson ρ (min-max) | Spearman ρ (min-max) |
|---|---|---|---|---|---|---|
| 1 | 10 | (54.5%-97.8%) | (4.3-19.2) | (19.0-70.5) | (0.416-0.956) | (0.596-0.962) |
| 2 | 8 | (80.0%-100.0%) | (16.8-29.0) | (47.2-100) | (0.130-0.737) | (0.204-0.721) |
| 3 | 11 | (77.8%-100.0%) | (7.2-39.3) | (17.6-93.3) | (-0.173-0.868) | (0.000-0.959) |
| 4 | 11 | (79.2%-100.0%) | (15.2-23.4) | (41.7-88.9) | (0.461-0.868) | (0.377-0.778) |
| 5 | 19 | (68.0%-100.0%) | (9.3-26.5) | (25.0-94.7) | (0.260-0.882) | (0.380-0.932) |
| 6 | 48 | (66.7%-100.0%) | (11.5-46.7) | (33.3-100) | (-0.005-0.929) | (-0.309-0.879) |
| 7 | 29 | (46.2%-100.0%) | (11.1-28.1) | (30.0-100) | (0.120-0.935) | (0.346-0.879) |
| 8 | 9 | (76.2%-93.5%) | (12.0-22.6) | (36.7-76.2) | (0.461-0.902) | (0.512-0.827) |
| 9 | 34 | (0.0%-100.0%) | (0.0-30.5) | (0.0-87.5) | (0.237-0.950) | (0.303-1.000) |

* 1 = Mathematics and computer sciences; 2 = Physics; 3 = Chemistry; 4 = Earth sciences; 5 = Biology; 6 = Medicine; 7 = Agricultural and veterinary sciences; 8 = Civil engineering; 9 = Industrial and information engineering.
** We excluded SDSs with less than 5 universities to be ranked

## 3.2 Comparing university scores and rankings at the discipline and overall level

For this analysis, we have measured the performance rankings by FSS and MNCS of all universities in each UDA. For the comparison by UDA, we exclude those universities with less than 10 professors in the UDA. To exemplify, Table 5 shows how the national rank in each UDA of a university (UNIV_1) changes when measuring performance by FSS and MNCS. We observe that this university does not change rank in Chemistry (UDA 3) and Agricultural and veterinary sciences (UDA 7), gains 15 positions in Physics and in Civil engineering, and 16 in Agricultural and veterinary sciences, and loses 15 in Mathematics. Overall, the university percentile rank is 66.7 by FSS and 88.9 by MNCS.

*Table 5: FSS and MNCS scores and related national ranks in a large generalist university (UNIV 1) per UDA*

| UDA* | Prof. | FSS | | | MNSC | | | Rank shift | Percentile shift |
|---|---|---|---|---|---|---|---|---|---|
| | | score | rank | percent. | score | rank† | percent. | | |
| 1 | 127 | 0.764 | 25 out of 49 | 50.0 | 0.581 | 40 out of 49 | 18.8 | ↓15 | -31.3 |
| 2 | 96 | 0.593 | 38 out of 43 | 11.9 | 0.672 | 23 out of 43 | 47.6 | ↑15 | +35.7 |
| 3 | 138 | 1.830 | 1 out of 44 | 100.0 | 1.320 | 1 out of 44 | 100.0 | = | 0.0 |
| 4 | 50 | 1.465 | 3 out of 32 | 93.5 | 0.987 | 5 out of 32 | 87.1 | ↓2 | -6.5 |
| 5 | 175 | 1.096 | 13 out of 53 | 76.9 | 0.899 | 15 out of 53 | 73.1 | ↓2 | -3.8 |
| 6 | 368 | 1.088 | 11 out of 42 | 75.6 | 0.848 | 8 out of 42 | 82.9 | ↑3 | +7.3 |
| 7 | 149 | 0.544 | 26 out of 29 | 10.7 | 0.775 | 10 out of 29 | 67.9 | ↑16 | +57.1 |
| 8 | 55 | 0.418 | 30 out of 36 | 17.1 | 0.711 | 15 out of 36 | 60.0 | ↑15 | +42.9 |
| 9 | 124 | 0.574 | 42 out of 47 | 10.9 | 0.532 | 42 out of 47 | 10.9 | = | 0.0 |
| Total | 1,282 | 0.973 | 22 out of 64 | 66.7 | 0.857 | 8 out of 64 | 88.9 | ↑14 | +22.2 |

*\* 1 = Mathematics and computer sciences; 2 = Physics; 3 = Chemistry; 4 = Earth sciences; 5 = Biology; 6 = Medicine; 7 = Agricultural and veterinary sciences; 8 = Civil engineering; 9 = Industrial and information engineering.*

Table 6 presents the comparison of university rankings in Chemistry (UDA 3). In this UDA, 44 universities have more than 10 professors. Figure 3 presents a graphic view of the dispersion of FSS and MNCS scores for these 44 universities: the correlation between the scores by the two indicators (Pearson ρ) is 0.504; the rank correlation (Spearman ρ) is 0.851.

To better appreciate the entity of the shifts and relevant frequencies, Figure 4 shows the frequency distribution of the percentile differences in rank. We observe that for about 11% of the universities (5 out of 44) the rank does not change, but for 16% of them (7 out of 44) the percentile rank shift is over 20 in absolute values.

*Table 6: FSS and MNCS scores and related rankings of Italian universities in Chemistry*

| ID* | Professors | FSS | | | MNCS | | | Rank shift | Percentile shift |
|---|---|---|---|---|---|---|---|---|---|
| | | score | rank | percentile | score | rank | percentile | | |
| UNIV_1 | 138 | 1.830 | 1 | 100.0 | 1.320 | 1 | 100.0 | = | 0.0 |
| UNIV_2 | 10 | 1.592 | 2 | 97.7 | 1.109 | 6 | 88.4 | ↓4 | -9.3 |
| UNIV_47 | 43 | 1.362 | 3 | 95.3 | 1.287 | 3 | 95.3 | = | 0.0 |
| UNIV_59 | 40 | 1.285 | 4 | 93.0 | 0.933 | 18 | 60.5 | ↓14 | -32.6 |
| UNIV_24 | 60 | 1.217 | 5 | 90.7 | 0.918 | 22 | 51.2 | ↓17 | -39.5 |
| UNIV_64 | 30 | 1.209 | 6 | 88.4 | 1.315 | 2 | 97.7 | ↑4 | +9.3 |
| UNIV_3 | 242 | 1.147 | 7 | 86.0 | 0.998 | 10 | 79.1 | ↓3 | -7.0 |
| UNIV_28 | 132 | 1.141 | 8 | 83.7 | 1.008 | 9 | 81.4 | ↓1 | -2.3 |
| UNIV_16 | 74 | 1.094 | 9 | 81.4 | 0.940 | 17 | 62.8 | ↓8 | -18.6 |
| UNIV_21 | 32 | 1.074 | 10 | 79.1 | 1.206 | 5 | 90.7 | ↑5 | +11.6 |
| UNIV_6 | 91 | 1.070 | 11 | 76.7 | 0.976 | 14 | 69.8 | ↓3 | -7.0 |
| UNIV_70 | 15 | 1.058 | 12 | 74.4 | 0.981 | 13 | 72.1 | ↓1 | -2.3 |
| UNIV_10 | 105 | 1.048 | 13 | 72.1 | 0.998 | 11 | 76.7 | ↑2 | +4.7 |
| UNIV_22 | 55 | 0.988 | 14 | 69.8 | 1.066 | 7 | 86.0 | ↑7 | +16.3 |
| UNIV_9 | 133 | 0.985 | 15 | 67.4 | 0.985 | 12 | 74.4 | ↑3 | +7.0 |
| UNIV_5 | 60 | 0.982 | 16 | 65.1 | 1.018 | 8 | 83.7 | ↑8 | +18.6 |
| UNIV_56 | 30 | 0.968 | 17 | 62.8 | 0.923 | 19 | 58.1 | ↓2 | -4.7 |
| UNIV_14 | 82 | 0.965 | 18 | 60.5 | 0.883 | 30 | 32.6 | ↓12 | -27.9 |
| UNIV_12 | 95 | 0.934 | 19 | 58.1 | 0.907 | 27 | 39.5 | ↓8 | -18.6 |
| UNIV_8 | 94 | 0.917 | 20 | 55.8 | 0.954 | 16 | 65.1 | ↑4 | +9.3 |
| UNIV_11 | 186 | 0.906 | 21 | 53.5 | 0.961 | 15 | 67.4 | ↑6 | +14.0 |
| UNIV_57 | 17 | 0.904 | 22 | 51.2 | 1.218 | 4 | 93.0 | ↑18 | +41.9 |
| UNIV_18 | 84 | 0.897 | 23 | 48.8 | 0.840 | 33 | 25.6 | ↓10 | -23.3 |
| UNIV_7 | 40 | 0.875 | 24 | 46.5 | 0.911 | 25 | 44.2 | ↓1 | -2.3 |
| UNIV_13 | 169 | 0.861 | 25 | 44.2 | 0.923 | 20 | 55.8 | ↑5 | +11.6 |
| UNIV_15 | 115 | 0.830 | 26 | 41.9 | 0.920 | 21 | 53.5 | ↑5 | +11.6 |
| UNIV_17 | 182 | 0.829 | 27 | 39.5 | 0.851 | 32 | 27.9 | ↓5 | -11.6 |
| UNIV_4 | 61 | 0.820 | 28 | 37.2 | 0.685 | 41 | 7.0 | ↓13 | -30.2 |
| UNIV_20 | 90 | 0.813 | 29 | 34.9 | 0.916 | 24 | 46.5 | ↑5 | +11.6 |
| UNIV_58 | 17 | 0.797 | 30 | 32.6 | 0.916 | 23 | 48.8 | ↑7 | +16.3 |
| UNIV_25 | 51 | 0.787 | 31 | 30.2 | 0.910 | 26 | 41.9 | ↑5 | +11.6 |
| UNIV_27 | 103 | 0.736 | 32 | 27.9 | 0.866 | 31 | 30.2 | ↑1 | +2.3 |
| UNIV_36 | 10 | 0.724 | 33 | 25.6 | 0.751 | 36 | 18.6 | ↓3 | -7.0 |
| UNIV_19 | 75 | 0.693 | 34 | 23.3 | 0.784 | 34 | 23.3 | = | 0.0 |
| UNIV_23 | 28 | 0.691 | 35 | 20.9 | 0.699 | 38 | 14.0 | ↓3 | -7.0 |
| UNIV_29 | 104 | 0.686 | 36 | 18.6 | 0.778 | 35 | 20.9 | ↑1 | +2.3 |
| UNIV_39 | 16 | 0.673 | 37 | 16.3 | 0.904 | 29 | 34.9 | ↑8 | +18.6 |
| UNIV_41 | 65 | 0.605 | 38 | 14.0 | 0.691 | 40 | 9.3 | ↓2 | -4.7 |
| UNIV_26 | 52 | 0.583 | 39 | 11.6 | 0.697 | 39 | 11.6 | = | 0.0 |
| UNIV_43 | 31 | 0.510 | 40 | 9.3 | 0.556 | 43 | 2.3 | ↓3 | -7.0 |
| UNIV_55 | 12 | 0.491 | 41 | 7.0 | 0.662 | 42 | 4.7 | ↓1 | -2.3 |
| UNIV_61 | 11 | 0.483 | 42 | 4.7 | 0.905 | 28 | 37.2 | ↑14 | +32.6 |
| UNIV_71 | 10 | 0.364 | 43 | 2.3 | 0.737 | 37 | 16.3 | ↑6 | +14.0 |
| UNIV_32 | 12 | 0.247 | 44 | 0.0 | 0.393 | 44 | 0.0 | = | 0.0 |

*\* The population consists of universities (44 in all) having at least 10 professors in the UDA*

*Figure 4: FSS and MNCS scores for Italian universities in Chemistry*

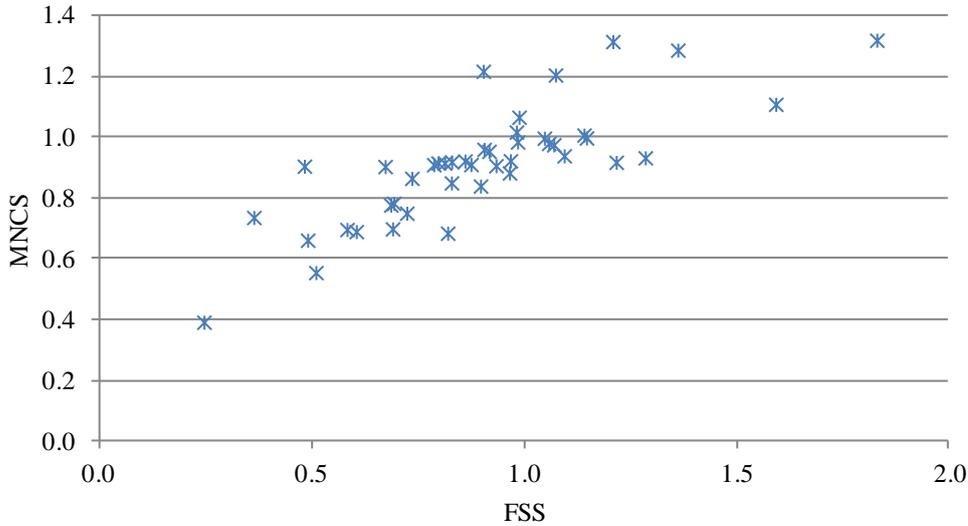

*Figure 5: Frequency distribution of university percentile rank shifts (by FSS and MNCS) in Chemistry (44 observations)*

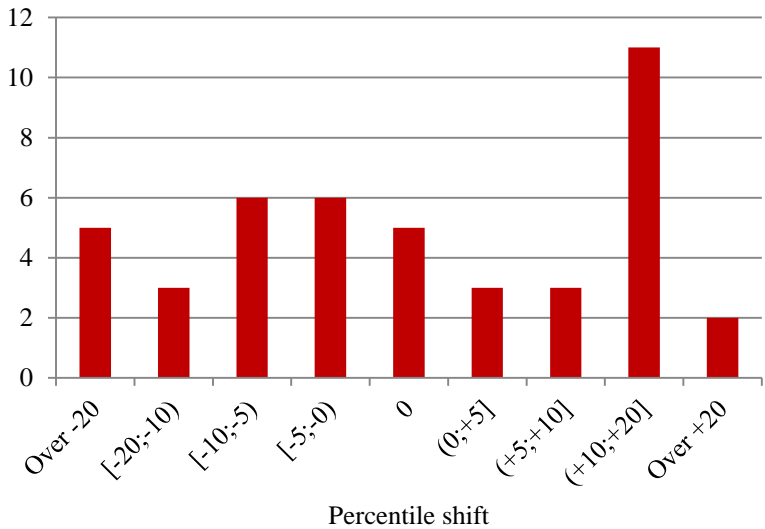

In Table 7 we present the descriptive statistics of university ranking shifts by FSS and MNCS in all UDAs. We observe that the number of universities experiencing rank shifts is never below 88.6%. The maximum percentile shift in ranking is never below 41.9 and can be as high as 78.6. Nevertheless, the correlation between the two rankings is strong in all UDAs, especially in Chemistry. The Pearson $\rho$ correlation between scores is minimum in Agricultural and veterinary sciences (0.451), while the Spearman $\rho$ in Industrial and information engineering (0.454). These two UDAs show the highest values of the average and maximum percentile shifts between the two rankings.

*Table 7: Score and rank correlations, and descriptive statistics of university ranking shifts by FSS and MNCS, at UDA level (percentile shift in brackets)*

| UDA* | No. of universities§ | % shifting rank | Average shift | Median shift | Max shift | Pearson ρ correlation | Spearman ρ correlation |
|---|---|---|---|---|---|---|---|
| 1 | 49 | 98.0% | 9.8 (20.5) | 7 | 33 (68.8) | 0.687 | 0.615 |
| 2 | 43 | 97.7% | 8.4 (19.9) | 5 | 27 (64.3) | 0.676 | 0.578 |
| 3 | 44 | 88.6% | 5.2 (12.1) | 4 | 18 (41.9) | 0.805 | 0.851 |
| 4 | 32 | 93.8% | 4.8 (15.3) | 3 | 14 (45.2) | 0.785 | 0.776 |
| 5 | 53 | 94.3% | 10.2 (19.6) | 8 | 37 (71.2) | 0.701 | 0.619 |
| 6 | 42 | 92.9% | 6.3 (15.3) | 4 | 27 (65.9) | 0.631 | 0.738 |
| 7 | 29 | 96.6% | 6.6 (23.4) | 5 | 22 (78.6) | 0.451 | 0.488 |
| 8 | 36 | 94.4% | 7.2 (20.6) | 5 | 25 (71.4) | 0.621 | 0.583 |
| 9 | 47 | 95.7% | 10.9 (23.6) | 9 | 35 (76.1) | 0.667 | 0.454 |

* 1 = Mathematics and computer sciences; 2 = Physics; 3 = Chemistry; 4 = Earth sciences; 5 = Biology; 6 = Medicine; 7 = Agricultural and veterinary sciences; 8 = Civil engineering; 9 = Industrial and information engineering.
§ The population consists of universities having at least 10 professors in the UDA

At overall level (Table 8) all universities shift rank but UNIV_42 and UNIV_59. The average percentile rank shift is 20; the median is 16. The maximum percentile shift is 66.7 for UNIV_34, which loses 42 positions passing from the FSS to the MNCS ranking. In Figure 5 we can appreciate the dispersion of FSS and MNCS scores (Pearson ρ = 0.574; Spearman ρ = 0.615).

The analysis of the distributions of the scores shows that the coefficient of variation by MNCS always falls below that by FSS (Table 9), revealing that the MNCS is less capable than FSS to observe significant performance differences.

Another possible way to classify universities is by quartile performance rankings. In Table 10 we report the descriptive statistics of the quartile shifts at UDA and overall level. At UDA level, the universities undergoing a quartile rank shift are no less than 36.4% in Chemistry (UDA 3), and as high as 63.8% in Industrial and information engineering (UDA 9). Except for Chemistry and Earth sciences, in all UDAs at least one university shifts from Q1 to Q4, or vice versa (max quartile shift = 3). Universities that shift from top quartile position to non top are 55.6% in Civil engineering (UDA 8) and 50% in Agricultural and veterinary sciences (UDA 7) and in Industrial and information engineering (UDA 9). At the overall level, almost one third of Q1 universities by FSS are non top by MNCS.

*Table 8: FSS and MNCS scores and related rankings of Italian universities*

| ID* | Professors | FSS | | | MNCS | | | Rank shift | Percentile shift |
|---|---|---|---|---|---|---|---|---|---|
| | | score | rank | percentile | score | rank | percentile | | |
| UNIV_68 | 72 | 2.814 | 1 | 100.0 | 1.107 | 2 | 98.4 | ↓1 | -1.6 |
| UNIV_52 | 55 | 1.933 | 2 | 98.4 | 0.809 | 15 | 77.8 | ↓13 | -20.6 |
| UNIV_63 | 43 | 1.852 | 3 | 96.8 | 1.120 | 1 | 100.0 | ↑2 | +3.2 |
| UNIV_66 | 62 | 1.646 | 4 | 95.2 | 1.008 | 3 | 96.8 | ↑1 | +1.6 |
| UNIV_62 | 85 | 1.454 | 5 | 93.7 | 0.696 | 40 | 38.1 | ↓35 | -55.6 |
| UNIV_35 | 232 | 1.422 | 6 | 92.1 | 0.801 | 16 | 76.2 | ↓10 | -15.9 |
| UNIV_42 | 398 | 1.225 | 7 | 90.5 | 0.881 | 7 | 90.5 | = | 0.0 |
| UNIV_36 | 163 | 1.216 | 8 | 88.9 | 0.884 | 6 | 92.1 | ↑2 | +3.2 |
| UNIV_9 | 1546 | 1.168 | 9 | 87.3 | 0.812 | 13 | 81.0 | ↓4 | -6.3 |
| UNIV_58 | 275 | 1.151 | 10 | 85.7 | 0.687 | 43 | 33.3 | ↓33 | -52.4 |
| UNIV_13 | 1608 | 1.116 | 11 | 84.1 | 0.810 | 14 | 79.4 | ↓3 | -4.8 |
| UNIV_64 | 955 | 1.115 | 12 | 82.5 | 0.763 | 24 | 63.5 | ↓12 | -19.0 |
| UNIV_34 | 170 | 1.103 | 13 | 81.0 | 0.652 | 55 | 14.3 | ↓42 | -66.7 |
| UNIV_28 | 1197 | 1.048 | 14 | 79.4 | 0.854 | 9 | 87.3 | ↑5 | +7.9 |
| UNIV_7 | 200 | 1.038 | 15 | 77.8 | 0.853 | 10 | 85.7 | ↑5 | +7.9 |
| UNIV_30 | 843 | 1.023 | 16 | 76.2 | 0.688 | 42 | 34.9 | ↓26 | -41.3 |
| UNIV_47 | 450 | 1.019 | 17 | 74.6 | 0.898 | 4 | 95.2 | ↑13 | +20.6 |
| UNIV_3 | 1807 | 1.018 | 18 | 73.0 | 0.787 | 19 | 71.4 | ↓1 | -1.6 |
| UNIV_56 | 271 | 1.010 | 19 | 71.4 | 0.722 | 31 | 52.4 | ↓12 | -19.0 |
| UNIV_24 | 480 | 1.005 | 20 | 69.8 | 0.726 | 30 | 54.0 | ↓10 | -15.9 |
| UNIV_38 | 114 | 0.974 | 21 | 68.3 | 0.643 | 57 | 11.1 | ↓36 | -57.1 |
| UNIV_1 | 1282 | 0.973 | 22 | 66.7 | 0.857 | 8 | 88.9 | ↑14 | +22.2 |
| UNIV_54 | 408 | 0.957 | 23 | 65.1 | 0.780 | 21 | 68.3 | ↑2 | +3.2 |
| UNIV_16 | 463 | 0.957 | 24 | 63.5 | 0.735 | 27 | 58.7 | ↓3 | -4.8 |
| UNIV_5 | 462 | 0.950 | 25 | 61.9 | 0.794 | 17 | 74.6 | ↑8 | +12.7 |
| UNIV_70 | 690 | 0.947 | 26 | 60.3 | 0.652 | 54 | 15.9 | ↓28 | -44.4 |
| UNIV_32 | 436 | 0.902 | 27 | 58.7 | 0.681 | 46 | 28.6 | ↓19 | -30.2 |
| UNIV_39 | 427 | 0.896 | 28 | 57.1 | 0.721 | 32 | 50.8 | ↓4 | -6.3 |
| UNIV_61 | 286 | 0.876 | 29 | 55.6 | 0.718 | 34 | 47.6 | ↓5 | -7.9 |
| UNIV_10 | 1205 | 0.870 | 30 | 54.0 | 0.735 | 28 | 57.1 | ↑2 | +3.2 |
| UNIV_2 | 161 | 0.867 | 31 | 52.4 | 0.818 | 12 | 82.5 | ↑19 | +30.2 |
| UNIV_46 | 92 | 0.865 | 32 | 50.8 | 0.675 | 50 | 22.2 | ↓18 | -28.6 |
| UNIV_4 | 567 | 0.843 | 33 | 49.2 | 0.678 | 49 | 23.8 | ↓16 | -25.4 |
| UNIV_14 | 726 | 0.838 | 34 | 47.6 | 0.843 | 11 | 84.1 | ↑23 | +36.5 |
| UNIV_51 | 43 | 0.832 | 35 | 46.0 | 0.763 | 25 | 61.9 | ↑10 | +15.9 |
| UNIV_44 | 121 | 0.822 | 36 | 44.4 | 0.625 | 59 | 7.9 | ↓23 | -36.5 |
| UNIV_59 | 1026 | 0.821 | 37 | 42.9 | 0.714 | 37 | 42.9 | = | 0.0 |
| UNIV_23 | 348 | 0.821 | 38 | 41.3 | 0.719 | 33 | 49.2 | ↑5 | +7.9 |
| UNIV_19 | 607 | 0.817 | 39 | 39.7 | 0.717 | 35 | 46.0 | ↑4 | +6.3 |
| UNIV_21 | 153 | 0.813 | 40 | 38.1 | 0.886 | 5 | 93.7 | ↑35 | +55.6 |
| UNIV_6 | 833 | 0.811 | 41 | 36.5 | 0.785 | 20 | 69.8 | ↑21 | +33.3 |
| UNIV_12 | 747 | 0.778 | 42 | 34.9 | 0.767 | 23 | 65.1 | ↑19 | +30.2 |
| UNIV_45 | 149 | 0.745 | 43 | 33.3 | 0.668 | 52 | 19.0 | ↓9 | -14.3 |
| UNIV_17 | 1980 | 0.743 | 44 | 31.7 | 0.735 | 29 | 55.6 | ↑15 | +23.8 |
| UNIV_15 | 1103 | 0.741 | 45 | 30.2 | 0.738 | 26 | 60.3 | ↑19 | +30.2 |
| UNIV_20 | 979 | 0.740 | 46 | 28.6 | 0.714 | 36 | 44.4 | ↑10 | +15.9 |
| UNIV_22 | 481 | 0.738 | 47 | 27.0 | 0.774 | 22 | 66.7 | ↑25 | +39.7 |
| UNIV_11 | 2750 | 0.714 | 48 | 25.4 | 0.697 | 39 | 39.7 | ↑9 | +14.3 |
| UNIV_55 | 702 | 0.714 | 49 | 23.8 | 0.710 | 38 | 41.3 | ↑11 | +17.5 |
| UNIV_72 | 81 | 0.668 | 50 | 22.2 | 0.536 | 63 | 1.6 | ↓13 | -20.6 |
| UNIV_41 | 124 | 0.653 | 51 | 20.6 | 0.633 | 58 | 9.5 | ↓7 | -11.1 |
| UNIV_29 | 1052 | 0.646 | 52 | 19.0 | 0.647 | 56 | 12.7 | ↓4 | -6.3 |
| UNIV_26 | 242 | 0.645 | 53 | 17.5 | 0.685 | 44 | 31.7 | ↑9 | +14.3 |
| UNIV_27 | 1208 | 0.642 | 54 | 15.9 | 0.679 | 48 | 25.4 | ↑6 | +9.5 |
| UNIV_31 | 127 | 0.622 | 55 | 14.3 | 0.794 | 18 | 73.0 | ↑37 | +58.7 |

| | | | | | | | | |
|---|---|---|---|---|---|---|---|---|
| UNIV_8 | 897 | 0.614 | 56 | 12.7 | 0.682 | 45 | 30.2 | ↑11 | +17.5 |
| UNIV_71 | 273 | 0.600 | 57 | 11.1 | 0.620 | 61 | 4.8 | ↓4 | -6.3 |
| UNIV_18 | 697 | 0.591 | 58 | 9.5 | 0.664 | 53 | 17.5 | ↑5 | +7.9 |
| UNIV_43 | 254 | 0.581 | 59 | 7.9 | 0.569 | 62 | 3.2 | ↓3 | -4.8 |
| UNIV_40 | 133 | 0.557 | 60 | 6.3 | 0.679 | 47 | 27.0 | ↑13 | +20.6 |
| UNIV_57 | 476 | 0.553 | 61 | 4.8 | 0.672 | 51 | 20.6 | ↑10 | +15.9 |
| UNIV_25 | 452 | 0.546 | 62 | 3.2 | 0.688 | 41 | 36.5 | ↑21 | +33.3 |
| UNIV_53 | 30 | 0.390 | 63 | 1.6 | 0.522 | 64 | 0.0 | ↓1 | -1.6 |
| UNIV_50 | 31 | 0.281 | 64 | 0.0 | 0.623 | 60 | 6.3 | ↑4 | +6.3 |

\* *The population consists of universities (64 in all) having at least 30 professors overall in the SDSs under investigation*

*Figure 6: FSS and MNCS scores for Italian universities*

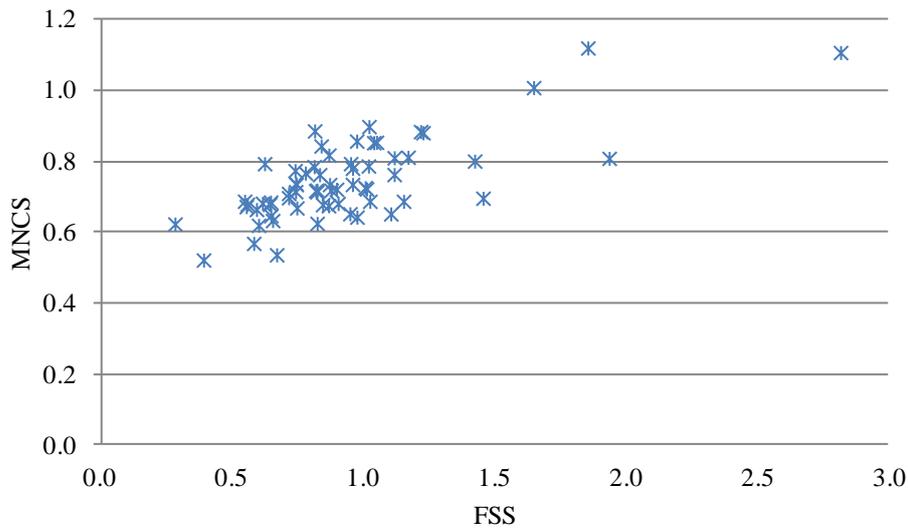

*Table 9: Descriptive statistics of FSS and MNCS score distributions, at UDA level*

| | FSS | | | MNCS | | |
|---|---|---|---|---|---|---|
| UDA | Average | Std dev. | Variation coeff. | Average | Std dev. | Variation coeff. |
| 1 | 0.830 | 0.335 | 0.404 | 0.693 | 0.181 | 0.260 |
| 2 | 0.970 | 0.515 | 0.531 | 0.722 | 0.174 | 0.241 |
| 3 | 0.902 | 0.300 | 0.333 | 0.914 | 0.188 | 0.206 |
| 4 | 0.844 | 0.410 | 0.486 | 0.789 | 0.182 | 0.231 |
| 5 | 0.939 | 0.350 | 0.373 | 0.836 | 0.156 | 0.187 |
| 6 | 0.920 | 0.430 | 0.467 | 0.751 | 0.144 | 0.191 |
| 7 | 0.835 | 0.285 | 0.342 | 0.742 | 0.175 | 0.236 |
| 8 | 0.719 | 0.338 | 0.470 | 0.692 | 0.171 | 0.247 |
| 9 | 0.892 | 0.336 | 0.376 | 0.628 | 0.112 | 0.179 |
| Total | 0.927 | 0.385 | 0.416 | 0.744 | 0.112 | 0.150 |

*Table 10: Descriptive statistics of university quartile rank shifts by FSS and MNCS, at UDA and overall level*

| UDA* | No. of universities§ | Shifting quartile | Average quartile shift | Max quartile shift | Shifting from Q1 |
|---|---|---|---|---|---|
| 1 | 49 | 61.2% | 0.8 | 3 | 30.8% |
| 2 | 43 | 48.8% | 0.7 | 3 | 36.4% |
| 3 | 44 | 36.4% | 0.4 | 1 | 36.4% |
| 4 | 32 | 43.8% | 0.5 | 2 | 37.5% |
| 5 | 53 | 58.5% | 0.8 | 3 | 42.9% |
| 6 | 42 | 47.6% | 0.6 | 3 | 36.4% |
| 7 | 29 | 55.2% | 0.8 | 3 | 50.0% |
| 8 | 36 | 61.1% | 0.8 | 3 | 55.6% |
| 9 | 47 | 63.8% | 0.9 | 3 | 50.0% |
| Total† | 64 | 48.4% | 0.7 | 3 | 31.3% |

\* 1 = Mathematics and computer sciences; 2 = Physics; 3 = Chemistry; 4 = Earth sciences; 5 = Biology; 6 = Medicine; 7 = Agricultural and veterinary sciences; 8 = Civil engineering; 9 = Industrial and information engineering.

§ The population consists of universities having at least 10 professors in the UDA

† The population consists of universities having overall at least 30 professors

## 5. Conclusions

Research performance rankings, whether commissioned or regularly published, do have an effect on the stakeholders of research systems. In some cases the rankings are specifically intended to inform research policies and strategic decisions at institution level, to allocate resources and incentivize researchers. In the case of the broadly published rankings, the ostensible, virtuous purposes include the intended reduction of asymmetric information between supply and demand for research and education, with the aim of permitting information-based choices and therefore market efficiency. From the popular response, it would appear that such rankings indeed reach a vast public, including influential stakeholders. Each recipient of the information embedded in performance rankings has different expectations and attributes different value to this information. Not least, the entities evaluated are themselves especially sensitive to the influence of their rank on their own reputation.

Assessment methodologies and indicators and their resulting ranking lists should first of all reflect the objective for which they have been constructed. However, most performance rankings, especially those published on a regular basis (Leiden, SCImago, Shanghai, and others following similar bibliometric models) address a generic audience, and we would therefore expect them to present little or no difference in their outcomes. Differences in ranks then reflect the different conceptual framework and operationalization of the measures used to build them.

In the conceptual works mentioned in the introduction (Abramo and D'Angelo, 2016a and 2016b), we argued strongly against the validity of the MNCS and all similar per-publication citation indicators as measures of research performance. We have refuted all institutional performance rankings based on them, and have urged the adoption of efficiency (output to input ratio) indicators, such as the FSS. However, we are aware that the availability of input data cannot be taken for granted. Given that input data can indeed be very difficult to access, the question has been put as to what extent institutional scores and rankings by MNCS are truly different from those by FSS. In this work we answer that question at field, discipline, and overall university level, showing the differences in scores and ranks. We have contrasted the Italian university scores and rankings by MNCS and FSS, at field, discipline, and overall university level. We have calculated the score and rankings correlation and the descriptive statistics of the shifts in rank at all levels. The correlations are strong in many cases but weak in others, especially at SDS level. A very high number of universities experience a shift in ranking both at SDS and UDA level, with extreme cases of maximum shift not passing unnoticed: at UDA level not less than one third of universities shift from top to non-top quartile ranks. At the overall university level ranking distributions seem correlated too but we registered 48.4% of universities shifting quartile and 31.3% top quartile universities by FSS positioned in non top quartiles by MNCS. Moreover, the FSS score distributions reveal a higher variation coefficient than the MNCS, certifying a lower capability of the latter to observe significant performance differences with respect to the former.

In the response to our above said critical article (Abramo and D'Angelo, 2016a), Bormann & Haunschild (2016) objected that to prove the superiority of a performance indicator over other ones, one should compare the relevant performance scores/rankings with a reference benchmark. If we took as a reference benchmark the opinion of all Italian academics, we would be most surprised if someone would agree with the MNCS ranking, for the following reason. In fact, we believe that in a number of countries, anybody would easily spot the top universities. In Italy, for example, no academics would object to the prestige and excellence of the School for Advanced Studies S. Anna in Pisa, or the University of Trento among public universities. Even the latest highly criticized Italian national research assessment exercise (VQR 2004-2010) positions them among the top ones in the performance ranking. They are so by FSS but not by MNCS: the School for Advanced Studies S. Anna in Pisa loses 13 positions out of 64 (-20.6 percentile), shifting from $2^{nd}$ to $15^{th}$; and the University of Trento loses 10 positions (-15.9 percentile), shifting from $6^{th}$ ($1^{st}$ among public universities) to $16^{th}$.

In the light of the findings of this work, we hope that scientometricians now have more information to assess the trade-off between the costs of acquiring input data and the benefits of more valid research performance measures. The same holds true for policy makers and the management of research institutions, in terms of the costs of making input data available and the benefits of more precise and reliable performance scores.

**Acknowledgement**

We thank Ludo Waltman for the inspiration that led to this work, and for his precious comments and insights on an earlier draft of the paper.

# References

Abramo, G., Cicero, T., D'Angelo, C.A. (2012b). Revisiting size effects in higher education research productivity. *Higher Education*, 63(6), 701-717.

Abramo, G., Cicero, T., D'Angelo, C.A. (2012c). Revisiting the scaling of citations for research assessment. *Journal of Informetrics*, 6(4), 470–479.

Abramo, G., D'Angelo, C.A. (2016a). A farewell to the MNCS and like size-independent indicators. *Journal of Informetrics*, 10(3), 646-651.

Abramo, G., D'Angelo, C.A. (2016b). A farewell to the MNCS and like size-independent indicators: Rejoinder. *Journal of Informetrics,* 10(3), 679-683.

Abramo, G., D'Angelo, C.A. (2015). Evaluating university research: same performance indicator, different rankings. *Journal of Informetrics*, 9(3), 514-525.

Abramo, G., D'Angelo, C.A. (2014). How do you define and measure research productivity? *Scientometrics,* 101(2), 1129-1144.

Abramo, G., D'Angelo, C.A. (2015a). Ranking research institutions by the number of highly-cited articles per scientist. *Journal of Informetrics,* 9(4), 915-923.

Abramo, G., D'Angelo, C.A. (2015b). Publication rates in 192 research fields. In A. Salah, Y. Tonta, A.A.A. Salah, C. Sugimoto (Eds) Proceedings of the *15$^{th}$ International Society of Scientometrics and Informetrics Conference - (ISSI - 2015)* (pp. 909-919). Istanbul: Bogazici University Printhouse. ISBN 978-975-518-381-7.

Abramo, G., D'Angelo, C.A., Cicero, T. (2012a). What is the appropriate length of the publication period over which to assess research performance? *Scientometrics*, 93(3), 1005-1017.

Abramo, G., D'Angelo, C.A., Di Costa, F. (2008b). Assessment of sectoral aggregation distortion in research productivity measurements. *Research Evaluation*, 17(2), 111-121.

Abramo, G., D'Angelo, C.A., Di Costa, F. (2010). Testing the trade-off between productivity and quality in research activities. *Journal of the American Society for Information Science and Technology*, 61(1), 132-140.

Abramo, G., D'Angelo, C.A., Solazzi, M. (2010). National research assessment exercises: a measure of the distortion of performance rankings when labor input is treated as uniform. *Scientometrics*, 84(3), 605-619.

Abramo, G., D'Angelo, C.A., Viel, F. (2013a). The suitability of h and g indexes for measuring the research performance of institutions. *Scientometrics*, 97(3), 555-570.

Abramo, G., D'Angelo, C.A., Viel, F. (2013b). Assessing the accuracy of the h- and g-indexes for measuring researchers' productivity. *Journal of the American Society for Information Science and Technology*, 64(6), 1224-1234.

Abramo, G., D'Angelo, C. A., Rosati, F. (2013c). Measuring institutional research productivity for the life sciences: the importance of accounting for the order of authors in the byline. *Scientometrics*, 97(3), 779-795.

Abramo, G., D'Angelo, C.A., Pugini, F. (2008a). The measurement of Italian universities' research productivity by a non parametric-bibliometric methodology. *Scientometrics*, 76 n. (2), 225-244.

Alonso, S., Cabrerizo, F.J., Herrera-Viedma, E., Herrera, F. (2009). h-Index: A review focused in its variants, computation and standardization for different scientific fields. *Journal of Informetrics*, 3(4), 273-289.

Bonaccorsi, A., Daraio, C. (2005). Exploring size and agglomeration effects on public research productivity. *Scientometrics*, 63(1), 87-120.

Bornmann, L., & Haunschild, R. (2016). Efficiency of research performance and the glass researcher. *Journal of Informetrics,* doi:10.1016/j.joi.2015.11.009.

Butler, L. (2007). Assessing university research: A plea for a balanced approach. *Science and Public Policy*, 34(8), 565-574.

D'Angelo, C.A., Giuffrida C., Abramo, G. (2011). A heuristic approach to author name disambiguation in bibliometrics databases for large-scale research assessments. *Journal of the American Society for Information Science and Technology*, 62(2), 257-269.

Egghe, L. (2010). The Hirsch index and related impact measures. *Annual Review of Information Science and Technology*, 44(1), 65-114.

Garfield, E. (1979). Is citation analysis a legitimate evaluation tool? *Scientometrics*, 1(4), 359-375.

Golden, J., Carstensen, F.V. (1992). Academic research productivity, department size and organization: Further results, comment. *Economics of Education Review,* 11(2), 169-171.

Hirsch, J.E. (2005). An index to quantify an individual's scientific research output. *Proceedings of the National Academy of Sciences of the United States of America*, 102(46), 16569-16572.

Larivière, V. & Costas, R., (2015). How many is too many? On the relationship between output and impact in research. In A. Salah, Y. Tonta, A.A.A. Salah, C. Sugimoto (Eds) Proceedings of the *15th International Society of Scientometrics and Informetrics Conference - (ISSI - 2015)* (pp. 590-595). Istanbul: Bogazici University Printhouse.

Moed H.F., De Bruin R.E., van Leeuwen Th.N., (1995). New bibliometric tools for assesment of national research performance: database description, overview of indicators and first applications. *Scientometrics* 33 (3), 381-422.

Moed, H.F., De Bruin, R.E., Van Leeuwen, Th. N. (1995). New bibliometric tools for the assessment of national research performance: Database description, overview of indicators and first applications. *Scientometrics*, 33(3), 381-422.

Norris, M., Oppenheim, C. (2010). The *h* index: a broad review of a new bibliometric indicator, *Journal of Documentation*, 66(5), 681-705.

Ruiz-Castillo, J., & Waltman, L. (2015). Field-normalized citation impact indicators using algorithmically constructed classification systems of science. *Journal of Informetrics, 9*(1), 102-117

Schubert, T. (2009). Empirical observations on New Public Management to increase efficiency in public research—Boon or bane? *Research Policy*, 38(8), 1225–1234.

Seglen, P.O., Asknes, D.G. (2000). Scientific productivity and group size: A bibliometric analysis of Norwegian microbiological research. *Scientometrics,* 49(1), 123-143.

van den Besselaar, P. & Sandström U. (2015). Does quantity make a difference? The importance of publishing many papers. In A. Salah, Y. Tonta, A.A.A. Salah, C. Sugimoto (Eds) Proceedings of the *15th International Society of Scientometrics and Informetrics Conference - (ISSI - 2015)* (pp. 577-583). Istanbul: Bogazici University Printhouse.

van Raan, A.F.J. (2005). Measuring science: Capita selecta of current main issues. In H.F. Moed, W. Glänzel, U. Schmoch (Eds.), *Handbook of quantitative science and*

*technology research* (pp. 19-50). Springer, ISBN 978-1-4020-2755-0.

Waltman, L., Van Eck, N.J., Van Leeuwen, T.N., Visser, M.S., Van Raan, A.F.J. (2011). Towards a new crown indicator: Some theoretical considerations. *Journal of Informetrics*, 5(1), 37-47.